\documentclass[10pt, twocolumn, pre, aps, superscriptaddress, showpacs]{revtex4-1}
\usepackage{amsmath, graphicx, subfigure, tikz}
\begin{document}

\title{Ferromagnetic and Spin-Glass Finite-Temperature Order\\
but no Antiferromagnetic Order in the d=1 Ising Model\\
with Long-Range Power-Law Interactions}

\author{E. Can Artun}
    \affiliation{TUBITAK Research Institute for Fundamental Sciences (TBAE), Gebze, Kocaeli 41470, Turkey}
    \affiliation{Faculty of Engineering and Natural Sciences, Kadir Has University, Cibali, Istanbul 34083, Turkey}
\author{A. Nihat Berker}
    \affiliation{Faculty of Engineering and Natural Sciences, Kadir Has University, Cibali, Istanbul 34083, Turkey}
    \affiliation{Department of Physics, Massachusetts Institute of Technology, Cambridge, Massachusetts 02139, USA}

\begin{abstract}
The $d=1$ Ising ferromagnet and spin glass with long-range power-law interactions $J\,r^{-a}$ is studied for all interaction range exponents $a$ by a renormalization-group transformation that simultaneously projects local ferromagnetism and antiferromagnetism.  In the ferromagnetic case, $J>0$, a finite-temperature ferromagnetic phase occurs for interaction range $0.74<a<2$. The second-order phase transition temperature monotonically decreases between these two limits.  At $a=2$, the phase transition becomes first order, also as predicted by rigorous results.  For $a>2$, the phase transition temperature discontinuously drops to zero and for $a>2$ there is no ordered phase above zero temperature, also as predicted by rigorous results. At the other end, on approaching $a=0.74$ from above, namely increasing the range of the interaction, the phase transition temperature diverges to infinity, meaning that, at all non-infinite temperatures, the system is ferromagnetically ordered.  Thus, the equivalent-neighbor interactions regime is entered before $(a > 0)$ the neighbors become equivalent, namely before the interactions become equal for all separations. The critical exponents $\alpha, \beta,\gamma,\delta,\eta,\nu$ are calculated, from a large recursion matrix, varying as a function of $a$. For the antiferromagnetic case, $J<0$, all triplets of spins at all ranges have competing interactions and this highly frustrated system does not have an ordered phase. In the spin-glass system, where all couplings for all separations are randomly ferromagnetic or antiferromagnetic (with probability $p$), a finite-temperatures spin-glass phase is obtained in the absence of antiferromagnetic phase.  A truly unusual phase diagram is obtained.  In the spin-glass phase, the signature chaotic behavior under scale change occurs in a richer version than previously:  In the long-range interaction of this system, the interactions at every separation become chaotic, yielding a piecewise chaotic interaction function.
\end{abstract}
\maketitle

\section{Ordering in One Dimension: Long-Range Interactions}

Whereas systems with finite-range interactions do not order above zero temperature in one dimension, certain systems with long-range interactions do order.\cite{Thouless,Ruelle,Griffiths,Aizenman1,Aizenman2} The archetypical example are the Ising ferromagnetic models with power-law interactions, $J\,r^{-a}$.  Also as seen below, for antiferromagnetic interactions, the system incorporates saturated frustration and spin-glass ordering without antiferromagnetic ordering, in the absence of quenched randomness.

The model that we study is defined by the Hamiltonian
\begin{equation}
- \beta {\cal H} = \sum_{r_1\neq r_2} \, J\,|r_1-r_2|^{-a} s_{r_1} s_{r_2}\, + \, H \sum_{r_1} \, s_{r_1} \,
\end{equation}
where $\beta = 1/k_B T$ is the inverse temperature, $r_1$ and $r_2$ designate the sites on the one-dimensional system, at each site $r_i$ there is an Ising spin $s_{r_i} = \pm 1$, and the sums are over all sites in the system.  For ferromagnetic and antiferromagnetic systems, the two-spin interactions $J$ are $J=|J| > 0$ and $J=-|J| < 0$, respectively.  For the spin-glass system, for each two spins at any range, their interaction is randomly ferromagnetic (with probability $1-p$) or antiferromagnetic (with probability $p$).  The second term in Eq. (1) is the magnetic-field $H$ term.

\begin{figure}[ht!]
\centering
\includegraphics[scale=0.6]{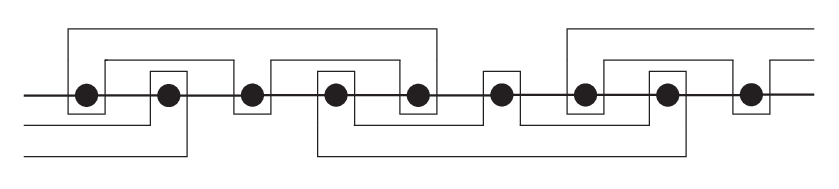}
\caption{Renormalization-group cells for $d=1$.  This cell structure projects both local ferromagnetism and antiferromagnetism, and therefore also spin-glass order.}
\end{figure}

\section{Method: Long-Range Renormalization Group}

We solve this system with Niemeyer and van Leeuwen's two-cell cluster approximation.\cite{NvL,vanLeeuwen,BerkerWor}  The renormalization-group transformation is constructed by first choosing cells on the $d=1$ system, as shown in Fig. 1.  Each of our cells has three spins. This cell structure projects both local ferromagnetism and antiferromagnetism, and therefore also spin-glass order. Secondly, for each cell, a cell-spin is defined as the sign of the sum of the three spins in the cell,
\begin{equation}
s'_{r'}= signum(s_{r-2}+s_{r}+s_{r+2})\,
\end{equation}
where the signum function returns the sign of its argument, primes denote the renormalized system, and $r'=r/b$, where $b=3$ is the length-rescaling factor.  The renormalized interactions are obtained from the conservation of the partition function $Z$,
\begin{multline}
Z=\sum_{\{s\}} e^{- \beta {\cal H}(\{s\})} = \sum_{\{s'\}} \sum_{\{\sigma\}} e^{- \beta {\cal H}  (\{s'\}  ,  \{\sigma\})} \\
 =\sum_{\{s'\}} e^{- \beta {\cal H}'(\{s'\})} =Z'\,,
\end{multline}
where the summed variable $\sigma$ represents, for each cell, the four states that give the same cell-spin value. Thus, the renormalized interactions are obtained from
\begin{equation}
e^{- \beta {\cal H}'(\{s'\})} =   \sum_{\{\sigma\}} e^{- \beta {\cal H}  (\{s'\}  ,  \{\sigma\})} .
\end{equation}

The two-cell cluster approximation of Niemeyer and van Leeuwen consists in carrying our this transformation for two cells, including the 6 intracell interactions and the 9 intercell interactions.  A recursion relation is obtained for each renormalized interaction,
\begin{multline}
J'_{r'} = \frac{1}{4} \ln \frac{R_{r'}(+1,+1)R_{r'}(-1,-1)}{R_{r'}(+1,-1)R_{r'}(-1,+1)}\,,\\
H' = \frac{1}{4} \ln \frac{R_{1}(+1,+1)}{R_{1}(-1,-1)}\,,
\end{multline}
where
\begin{equation}
R_{r'}(s'_0,s'_{r'}) =   \sum_{\sigma_0,\sigma_{r'}} e^{- \beta {\cal H}_{0r'}  } ,
\end{equation}
where the unrenormalized two-cell Hamiltonian contains the 6 intracell interactions and the 9 intercell interactions between the 6 spins in cells 0 and $r'$.

\begin{figure}[ht!]
\centering
\includegraphics[scale=0.43]{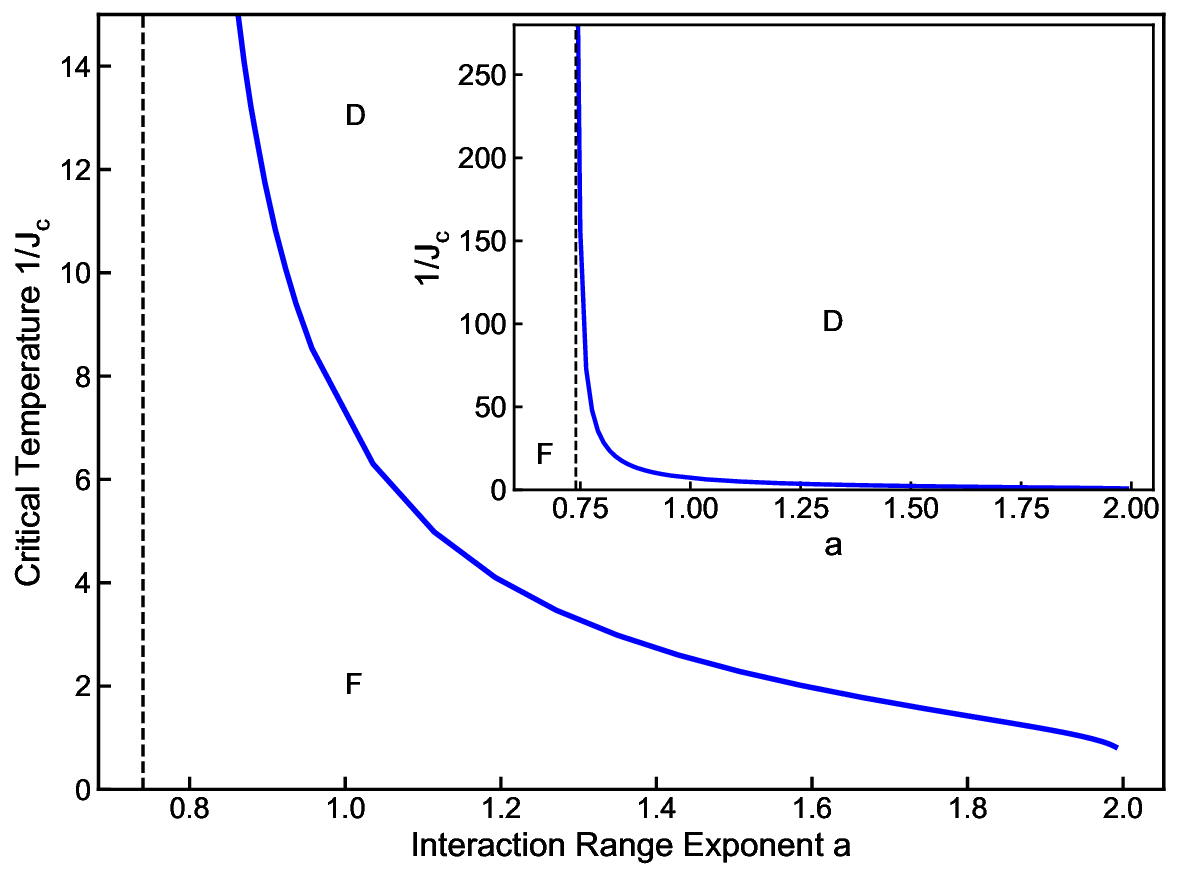}
\caption{Calculated phase diagram of the $d=1$ long-range ferromagnetic Ising model with interactions $J\,r^{-a}$. Ferromagnetic (F) and disordered (D) phases are seen. A finite-temperature ferromagnetic phase occurs for $0.74<a<2$. The second-order phase transition temperature monotonically decreases between these two limits.  At $a=2$, the transition becomes first-order, as predicted by rigorous results \cite{Aizenman2}.  For $a>2$ the phase transition temperature discontinuously drops to zero and there is no ordered phase above zero temperature, also as predicted by rigorous results \cite{Ruelle,Griffiths}.  At the other end, on approaching $a=0.74$ from above, the phase transition temperature diverges to infinity, meaning that, at all non-infinite temperatures, the system is ferromagnetically ordered.  Thus, the equivalent-neighbor interactions regime is entered before $(a > 0)$ the neighbors become equivalent, namely before the interactions become equal for all separations. To the left of the dashed line on this figure is the equivalent-neighbor regime.}
\end{figure}

\begin{figure*}[ht!]
\centering
\includegraphics[scale=0.47]{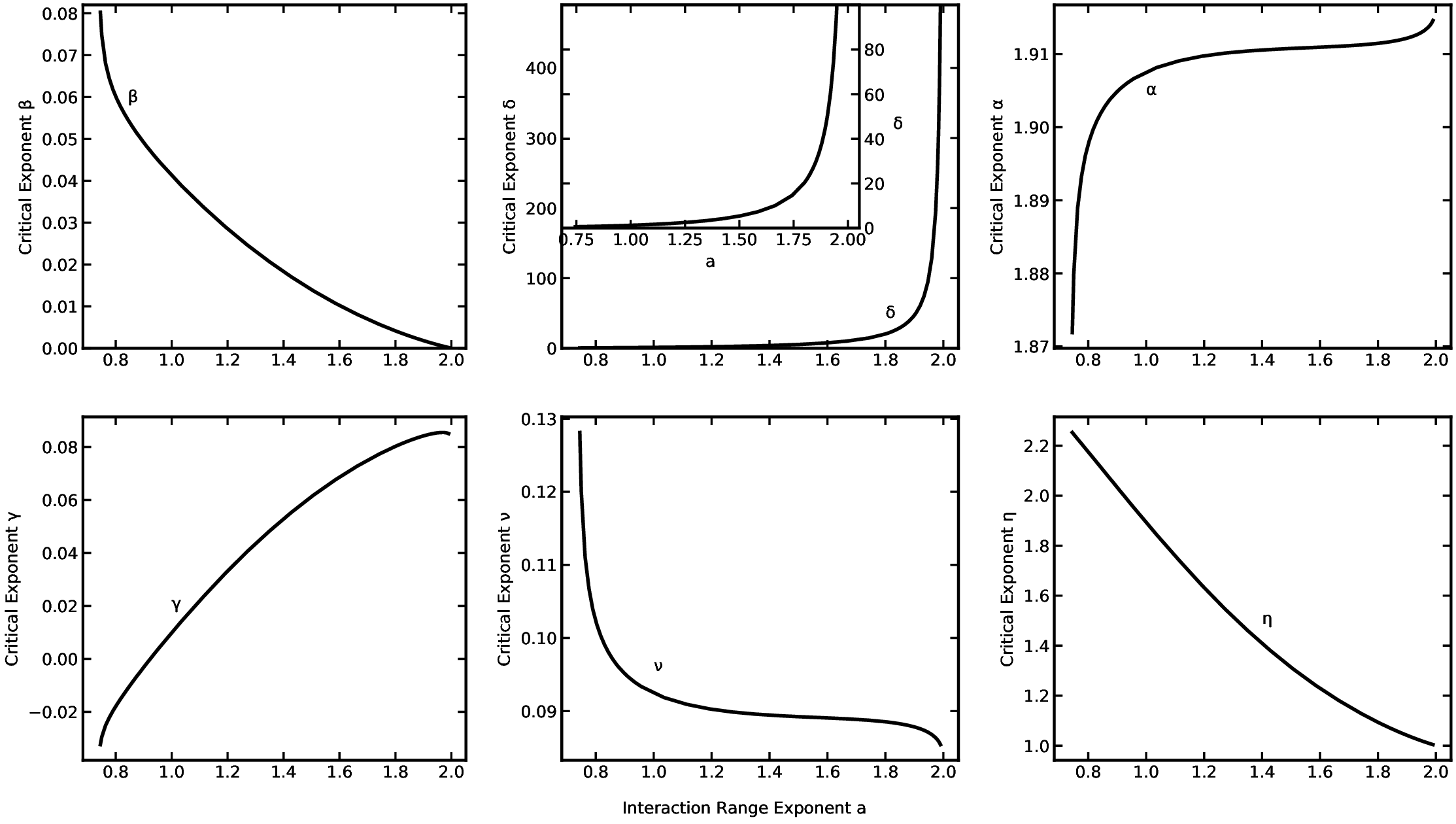}
\caption{Correlation-length critical exponent $\nu$, correlation-function critical exponent $\eta$, specific heat critical exponent $\alpha$, magnetization critical exponents $\beta$ and $\delta$, susceptibility critical exponent $\gamma$, as a function of interaction range $a$ for the finite-temperature ferromagnetic phase transition.  Note that $\beta$ reaches 0 and $\delta$ diverges to infinity, as expected, as the first-order phase transition as $a=2$ is reached from below.}
\end{figure*}

\section{Results: Finite-Temperature Ferromagnetic Phase in $d=1$}

The calculated phase diagram of the $d=1$ long-range ferromagnetic Ising model, with interactions $J\,r^{-a}$ , is shown in Fig. 2, in terms of  temperature $1/J$ and interaction range $a$. A finite-temperature ferromagnetic phase occurs for $0.74<a<2$. The second-order phase transition temperature monotonically decreases between these two limits.  At $a=2$, phase transition becomes first order, as predicted by rigorous results \cite{Aizenman2}.  For $a>2$, the phase transition temperature discontinuously drops to zero and there is no ordered phase above zero temperature, also as predicted by rigorous results.\cite{Ruelle,Griffiths}  At the other end, on approaching $a=0.74$ from above, the phase transition temperature diverges to infinity, meaning that, at all non-infinite temperatures, the system is ferromagnetically ordered.  Thus, the equivalent-neighbor interactions regime is entered before $(a > 0)$ the neighbors become equivalent, namely before the interactions become equal for all separations.

The calculated correlation-length critical exponent $\nu$, correlation-function critical exponent $\eta$, specific heat critical exponent $\alpha$, magnetization critical exponents $\beta$ and $\delta$, the susceptibility critical exponent $\gamma$, continuously varying as a function of interaction range $a$ for the finite-temperature ferromagnetic phase transition, are shown in Fig. 3. These critical exponents are calculated, with $H=H'=0$, from the relations $J'_1,...,J'_n = funct(J_1,...,J_n)$ of Eqs. (5,6). Convergence is obtained by calculation up to $n=20$. The largest (and, as expected, only relevant, namely greater than 1) eigenvalue $\lambda_T = b^{y_T}$ of the derivative matrix of these recursion relations at the critical point gives the correlation-length critical exponent $\nu = 1/y_T$ and the specific heat critical exponent $\alpha = 2-d/y_T = 2-1/y_T$. The magnetization critical exponents $\beta = (d-y_H)/y_T = (1-y_H)/y_T$ and $\delta = y_H/(d-y_H) = y_H/(1-y_H)$, the susceptibility critical exponent $\gamma = (2y_H-d)/y_T$, and the correlation-function critical exponent $\eta = 2+d -y_H = 3 -y_H$ are calculated, at the critical point, with $H=H'=0$, from $\partial H'/\partial H = b^{y_H}$.\cite{BerkerWor} Note that at $a=2$, the magnetization critical exponent $\beta=0$, which gives a first-order phase transition \cite{FisherBerker} as the temperature is scanned.  At $a=2$, the other magnetization critical exponent $\delta$ diverges to infitinity, which gives the first-transition as the magnetic field is scanned.

\begin{figure}[ht!]
\centering
\includegraphics[scale=0.48]{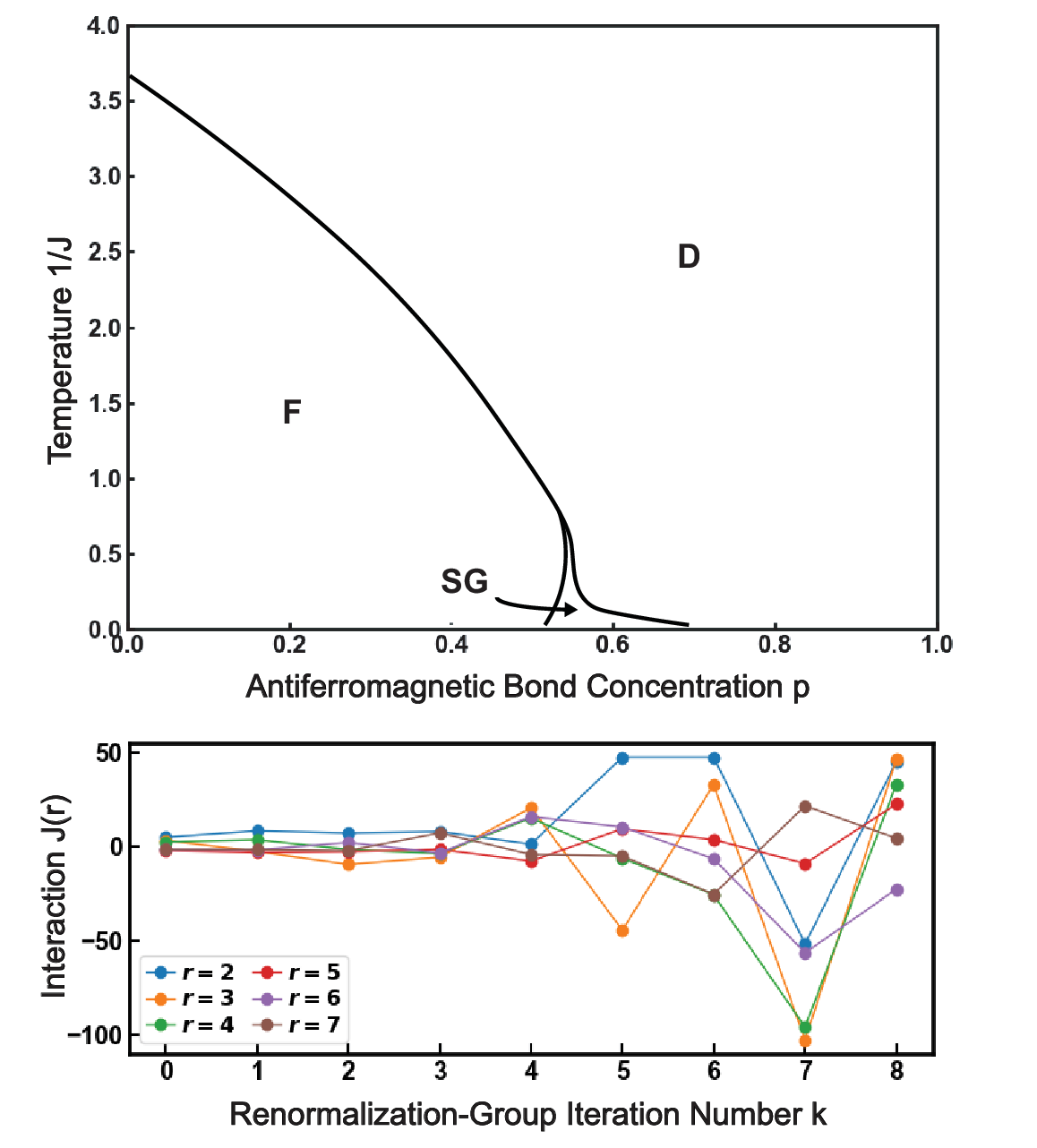}
\caption{Calculated finite-temperature phase diagram of the $d=1$ long-range Ising spin-glass system with interaction-range exponent $a=1$, where all couplings for all separations are randomly ferromagnetic or antiferromagnetic (with probability $p$).  Ferromagnetic (F), spin-glass (SG), and disordered (D) phases are seen.  This truly unusual spin-glass phase diagram, actually does not have an antiferromagnetic phase but has a spin-glass phase.  Bottom panel: Chaos inside the spin-glass phase in $d=1$. The spin-glass phase shows the chaos under rescaling signature \cite{McKayChaos,McKayChaos2,McKayChaos4,BerkerMcKay}, in a richer version than previously:  In the long-range interaction of this system, the interactions at every separation become chaotic, as seen in the lower panel of this figure, yielding a piecewise chaotic interaction potential.}
\end{figure}

The antiferromagnetic, overly frustrated without randomness, system does not have a finite-temperature phase transition, but the spin-glass system, where all couplings for all separations are randomly ferromagnetic or antiferromagnetic (with probability $p$), does have finite-temperature spin-glass phase transitions and chaos inside the spin-glass phase, as seen in Fig. 4.

\section{Results: Finite-Temperature Spin-Glass Phase in $d=1$}

The spin-glass system, where all couplings for all separations are randomly ferromagnetic or antiferromagnetic (with probability $p$), does have finite-temperature spin-glass phase transitions and chaos inside the spin-glass phase, as seen in Fig. 4.  This truly unusual spin-glass phase diagram, actually does not have an antiferromagnetic phase but has a spin-glass phase.  Nevertheless, typical spin-glass system reentrance \cite{ReentSG} is seen in this phase diagram, where as temperature is lowered at fixed antiferromagnetic bond concentration $p$, the ferromagnetic phase appears, but disappears at further lower temperature.

The spin-glass phase shows the chaos under rescaling signature \cite{McKayChaos,McKayChaos2,McKayChaos4,BerkerMcKay}, in a richer version than previously:  In the long-range interaction of this system, the interactions at every separation become chaotic, as seen in the lower panel of Fig. 4, yielding a piecewise chaotic interaction potential.

For a previous $d=1$ Ising spin-glass study, with short-range interactions and a zero-temperature spin-glass phase, see \cite{Grinstein}.

\section{Conclusion}

We have solved the $d=1$ Ising ferromagnet, antiferromagnet, and spin glass with long-range power-law interactions $J\,r^{-a}$, for all interaction range exponents $a$ by a renormalization-group transformation that simultaneously projects local ferromagnetism, antiferromagnetism, and spin-glass order.  In the ferromagnetic case, $J>0$, a finite-temperature second-order ferromagnetic phase occurs for interaction range $0.74<a<2$. The second-order phase transition temperature monotonically decreases between these two limits.  At $a=2$, the phase transition becomes first order, as predicted by rigorous results.  For $a>2$, the phase transition temperature discontinuously drops to zero and for $a>2$ there is no ordered phase above zero temperature, also as predicted by rigorous results. At the other end, on approaching $a=0.74$ from above, namely increasing the range of the interaction, the phase transition temperature diverges to infinity, meaning that, at all non-infinite temperatures, the system is ferromagnetically ordered.  Thus, the equivalent-neighbor interactions regime is entered before $(a > 0)$ the neighbors become equivalent, namely before the interactions become equal $(a = 0)$ for all separations. The critical exponents $\alpha, \beta,\gamma,\delta,\eta,\nu$ for the second-order phase transitions are calculated, from a large recursion matrix, varying as a function of $a$.

For the antiferromagnetic case, $J<0$, all triplets of spins at all ranges have competing interactions and this highly frustrated system does not have an ordered phase.

In the spin-glass system, where all couplings for all separations are randomly ferromagnetic or antiferromagnetic (with probability $p$), a finite-temperatures spin-glass phase is obtained in the absence of antiferromagnetic phase.  A truly unusual phase diagram, with reentrance around the ferromagnetic phase, is obtained.  In the spin-glass phase, the signature chaotic behavior under scale change occurs in a richer version than previously:  In the long-range interaction of this system, the interactions at every separation become chaotic, yielding a piecewise chaotic interaction function.

\begin{acknowledgments} Support by the Academy of Sciences of Turkey (T\"UBA) is gratefully acknowledged.
\end{acknowledgments}

\end{document}